\documentclass{moriond}

\def\be{\begin{equation}}
\def\ee{\end{equation}}
\def\bea{\begin{eqnarray}}
\def\eea{\end{eqnarray}}

\newcommand{\emu}{$e^{\pm}\mu^{\mp}$}
\newcommand{\pt}{p_{\mathrm{T}}}
\newcommand{\abseta}{|\eta|}
\newcommand{\GeV}{\,\mathrm{GeV}}
\newcommand{\DR}{\Delta R}
\newcommand{\ttbar}{$\mathrm{t}\bar{\mathrm{t}}$}

\newcommand{\Photo}{\includegraphics[height=35mm]{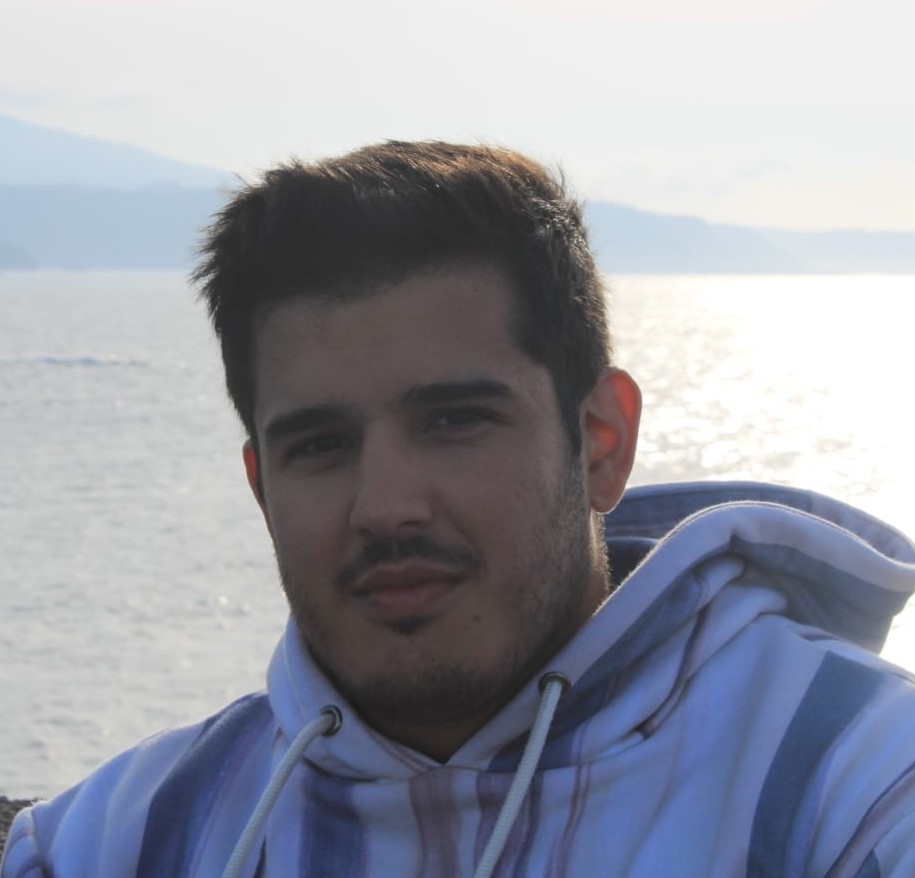}}

\begin{document}
\vspace*{4cm}
\title{Inclusive and differential cross-section measurements in tW with CMS in Run 3.}

\author{Alejandro Soto Rodr\'iguez on behalf of the CMS Collaboration}

\address{University of Oviedo - ICTEA, Spain}

\maketitle\abstracts{
   The first measurements of the inclusive and normalised differential cross sections for the production of single top quarks in association with a W boson in proton-proton collisions at a centre-of-mass energy of 13.6 TeV are presented. The data used were recorded with the CMS detector at the LHC during 2022, and correspond to an integrated luminosity of 34.7 $\mathrm{fb}^{-1}$. The analysed events contain one muon and one electron in the final state. A cross section of $84.1\pm 2.1(\mathrm{stat})\,^{+9.8}_{-10.2}(\mathrm{syst})\pm 3.3(\mathrm{lumi})$ pb is obtained, consistent with the predictions of the standard model. For the differential measurements, a fiducial region is defined according to the detector acceptance. The resulting differential distributions are unfolded to particle level showing good agreement with the predictions at next-to-leading order in perturbative quantum chromodynamics.
}

\section{Introduction}
Single top quarks were first observed by the D0~\cite{Abazov:2009ii} and CDF~\cite{Aaltonen:2009jj} Collaborations at the Fermilab Tevatron collider. The three main production modes of single top quarks in proton-proton (pp) collisions are mediated via electroweak interactions and are commonly categorised through the virtuality of the exchanged W boson four-momentum. When the four-momentum is space-like, the process is referred to as the $t$ channel, while when it is time-like, the process is referred to as the $s$ channel. The third production mode, referred to as the tW process, is characterised by the production of a top quark in association with an on-shell W boson.

The study of the tW process provides a unique opportunity to probe the standard model (SM) and its potential extensions through its interference with top quark pair production ($\mathrm{t}\bar{\mathrm{t}}$) at next-to-leading order (NLO) in quantum chromodynamics (QCD). In this analysis, two schemes are defined to avoid double counting issues: diagram removal (DR), in which all doubly resonant diagrams are removed from the matrix element (ME) calculation, and diagram subtraction (DS), in which a gauge invariant term is introduced in the ME calculation that locally cancels the doubly resonant diagrams. The tW production cross section is computed at approximate third-order (a$\mathrm{N}^3$LO) in QCD with the addition of third-order corrections of soft-gluon emission terms. The corresponding theoretical prediction for the tW cross section in pp collisions at $\sqrt{s} = 13.6$ TeV, assuming a top quark mass ($m_{\mathrm{t}}$) of 172.5 GeV~\cite{Kidonakis:2021vob,Kidonakis:2016sjf,Kidonakis:2010ux}, is 
\begin{equation}
    \sigma^{\mathrm{SM}}_{\mathrm{tW}} = 87.9 ^{+2.0}_{-1.9}\, (\mathrm{scale}) \pm 2.4\, (\mathrm{PDF}+\alpha_{\mathrm{S}}) \ \mathrm{pb.}
    \label{eq:Smtw}
\end{equation}

This document reports the first measurements from the CMS Collaboration of the inclusive and differential cross sections of the tW process in pp collisions at $\sqrt{s} = 13.6$ TeV~\cite{CMS:2024jmu}. The measurements are performed using final states with one electron and one muon of opposite charge. The analysed data was recorded by the CMS detector~\cite{Chatrchyan:2008zzk,CMS:2023gfb} during 2022 and corresponds to $34.7 \mathrm{ fb}^{-1}$ of integrated luminosity.

\section{Event selection} 
Events in which the W boson from the decay of the top quark and the W boson produced in association with the top quark both decay leptonically are used.
The events with same flavour leptons are rejected due to their high background contribution. 
This leads to a final state composed of two different-flavour leptons with opposite electric charge, \emu, one jet resulting from the fragmentation of a bottom quark, and two neutrinos. 

Electrons and muons in the event are required to have $\pt > 20\GeV$ and $\abseta < 2.4$. Jets are required to have $\pt > 30\GeV$ and $\abseta < 2.4$, and are required to be separated from any selected lepton by $\DR > 0.4$. Loose jets are defined as jets with $\pt$ between 20 and 30$\GeV$ and $\abseta < 2.4$.

\begin{figure}[htpb!]
\begin{center}
\begin{minipage}{0.328\linewidth}
\centerline{\includegraphics[width=1.0\linewidth]{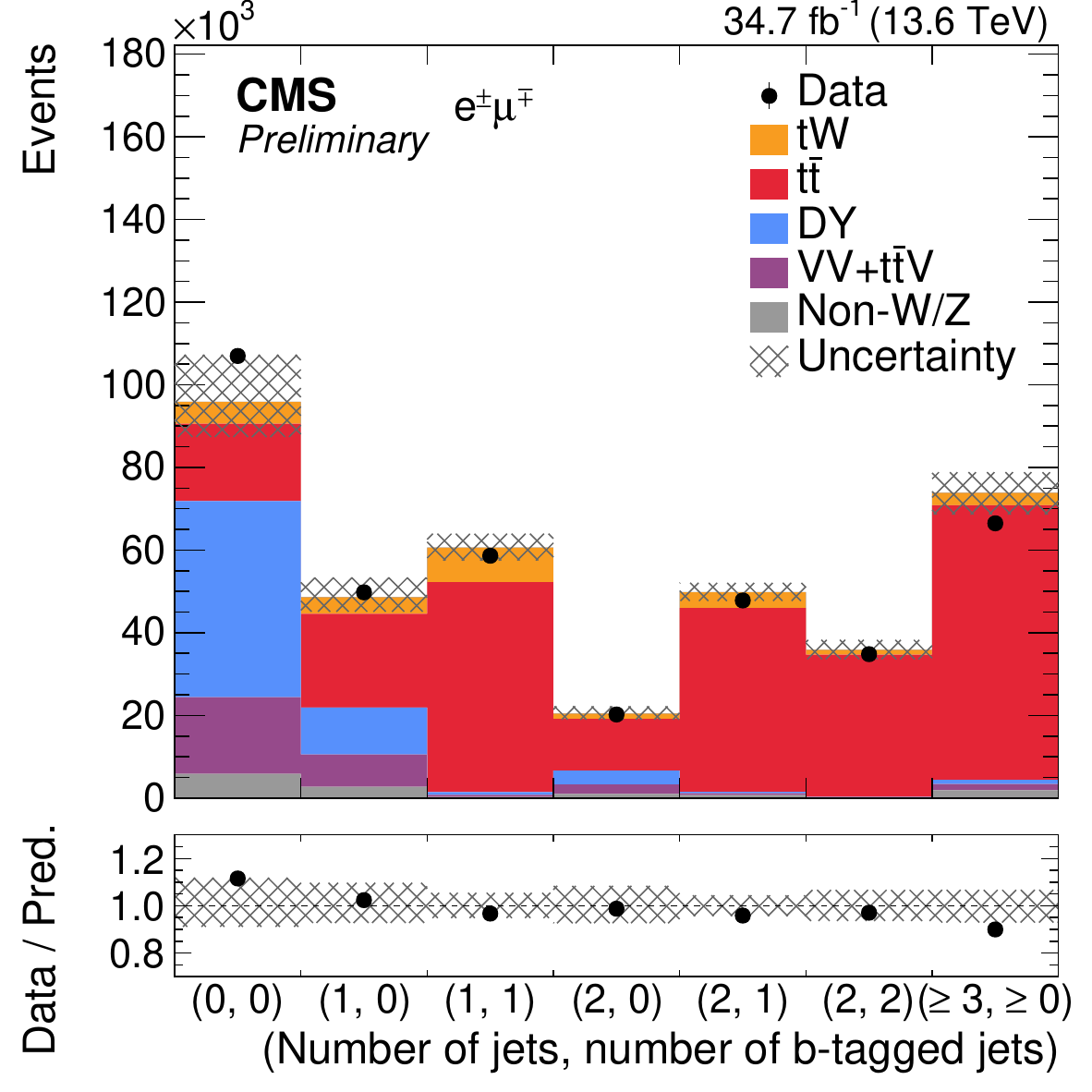}}
\end{minipage}
\begin{minipage}{0.328\linewidth}
\centerline{\includegraphics[width=1.0\linewidth]{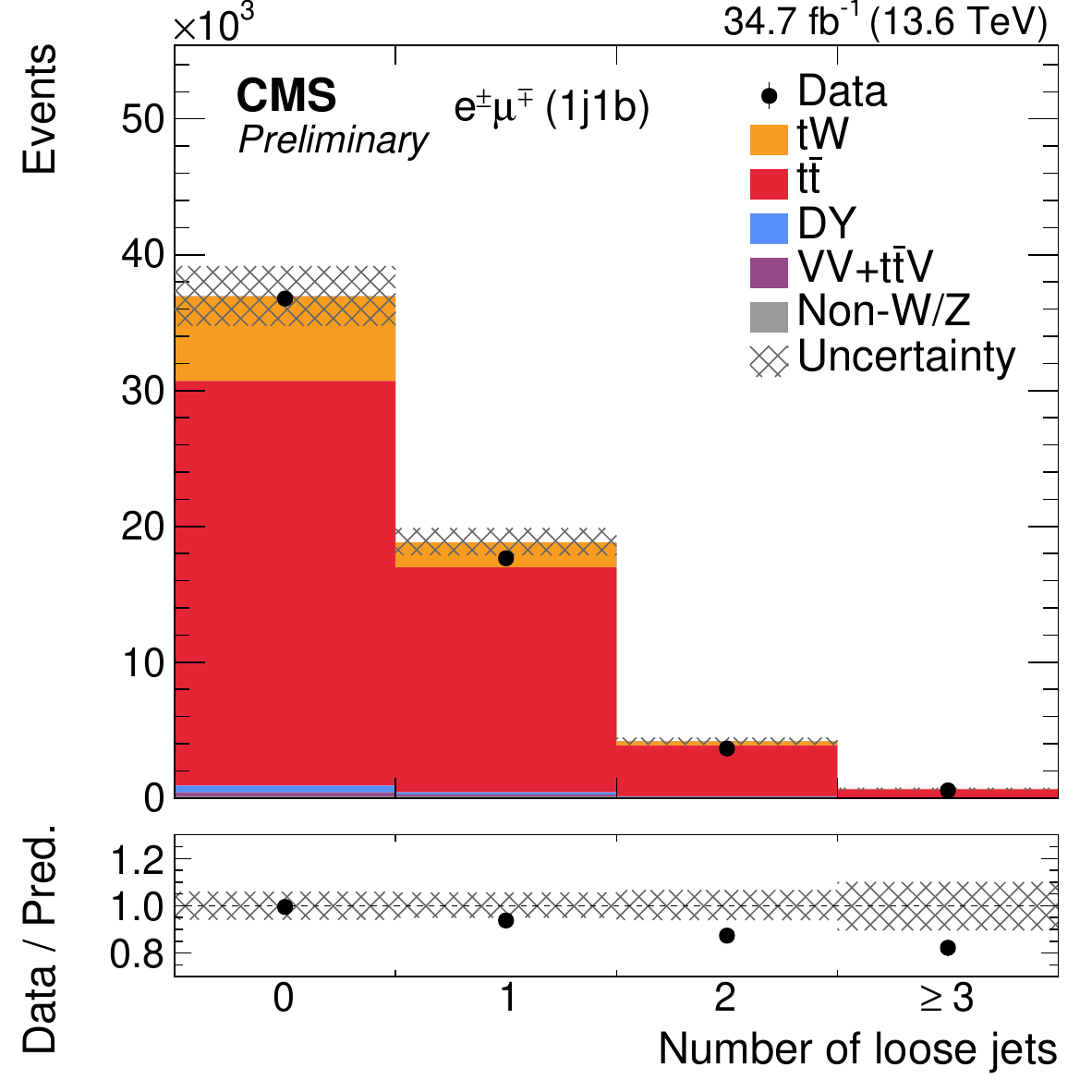}}
\end{minipage}
\end{center}
\caption[]{Number of events observed in data (points) and predicted from simulation (filled histograms) as a function of the number of jets and b-tagged jets for events passing the baseline dileptons selection (left) and number of loose jets in the 1j1b region (right)~\cite{CMS:2024jmu}.}
\label{fig:njnb}
\end{figure}

The selected events belong to the \emu $ $ final state if the two leptons with highest $\pt$ passing the above selection criteria  are an electron and a muon of opposite electric charge. The highest $\pt$ (leading) lepton is required to have $\pt$ greater than 25$\GeV$. In addition, to reduce the contamination from low-mass resonances, the minimum invariant mass of all pairs of identified leptons is required to be greater than 20$\GeV$. The remaining events are classified by the number of jets and the number of identified b jets in the event, as shown in Fig.~\ref{fig:njnb} (left). For the inclusive measurement the regions with one b-tagged jet (1j1b), two jets and one of them b-tagged (2j1b) and two b-tagged jets (2j2b) are used, while for the differential measurements only the 1j1b region is used. Figure~\ref{fig:njnb} (right) shows the distribution of the number of loose jets in the 1j1b region. In order to decrease the relative contribution from the $\mathrm{t}\bar{\mathrm{t}}$ background, the events in the 1j1b region with zero loose jets are used for the differential measurements.

\section{Inclusive cross section measurement}
One of the main challenges of this measurement is to distinguish the signal contribution from the dominant \ttbar $ $ background contribution. For this reason, two independent random forest (RF) multiclassifiers~\cite{randomforest}, one for the 1j1b region and the other for the 2j1b region, are trained to discriminate between \ttbar, tW and the second largest background in each category. For the 1j1b region, the second largest background is the Drell--Yan (DY) process, whereas for the 2j1b region, semileptonic \ttbar $ $ production is the one that is considered.

\begin{figure}[htpb!]
\begin{minipage}{0.328\linewidth}
\centerline{\includegraphics[width=\linewidth]{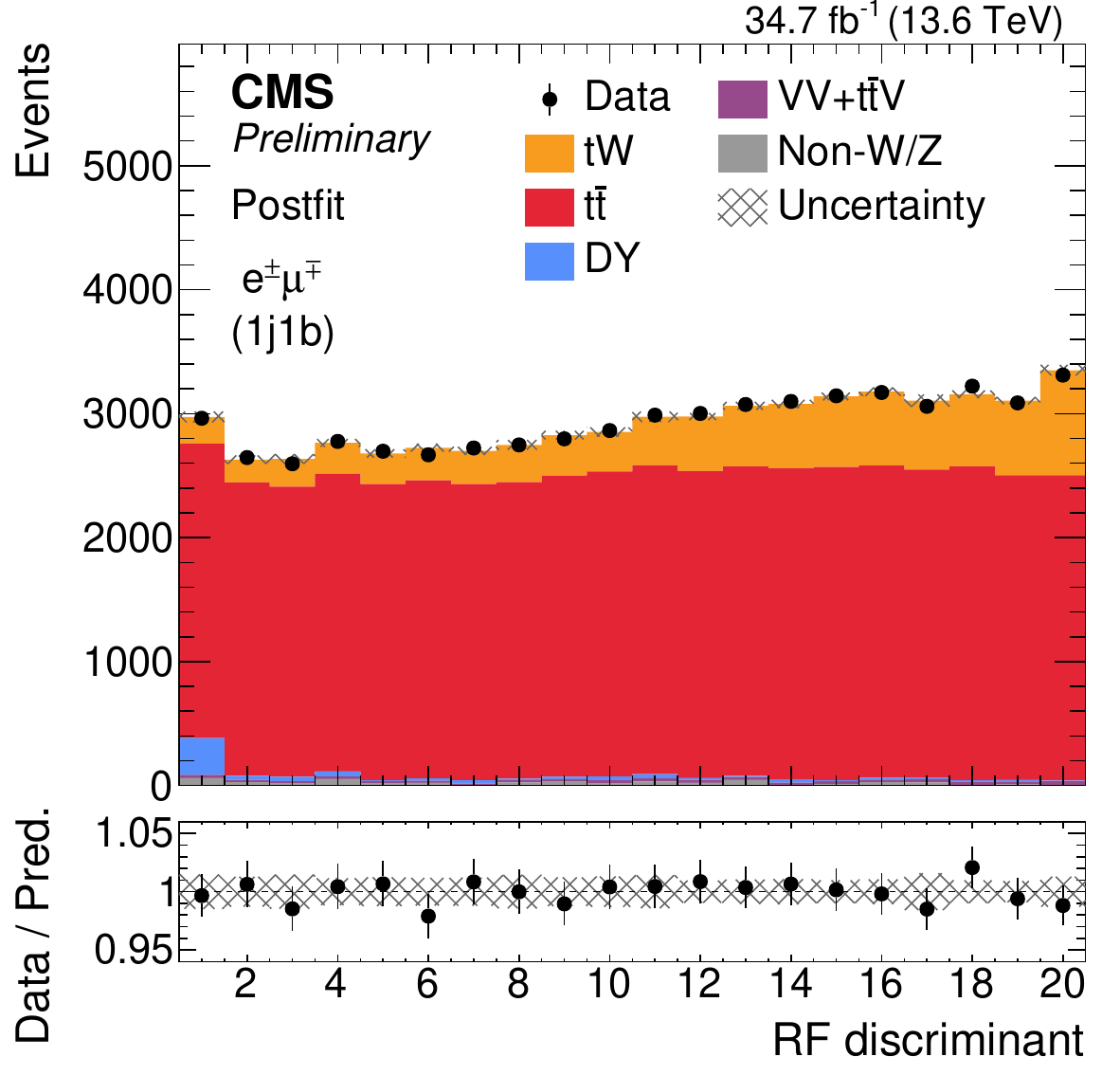}}
\end{minipage}
\hfill
\begin{minipage}{0.328\linewidth}
\centerline{\includegraphics[width=\linewidth]{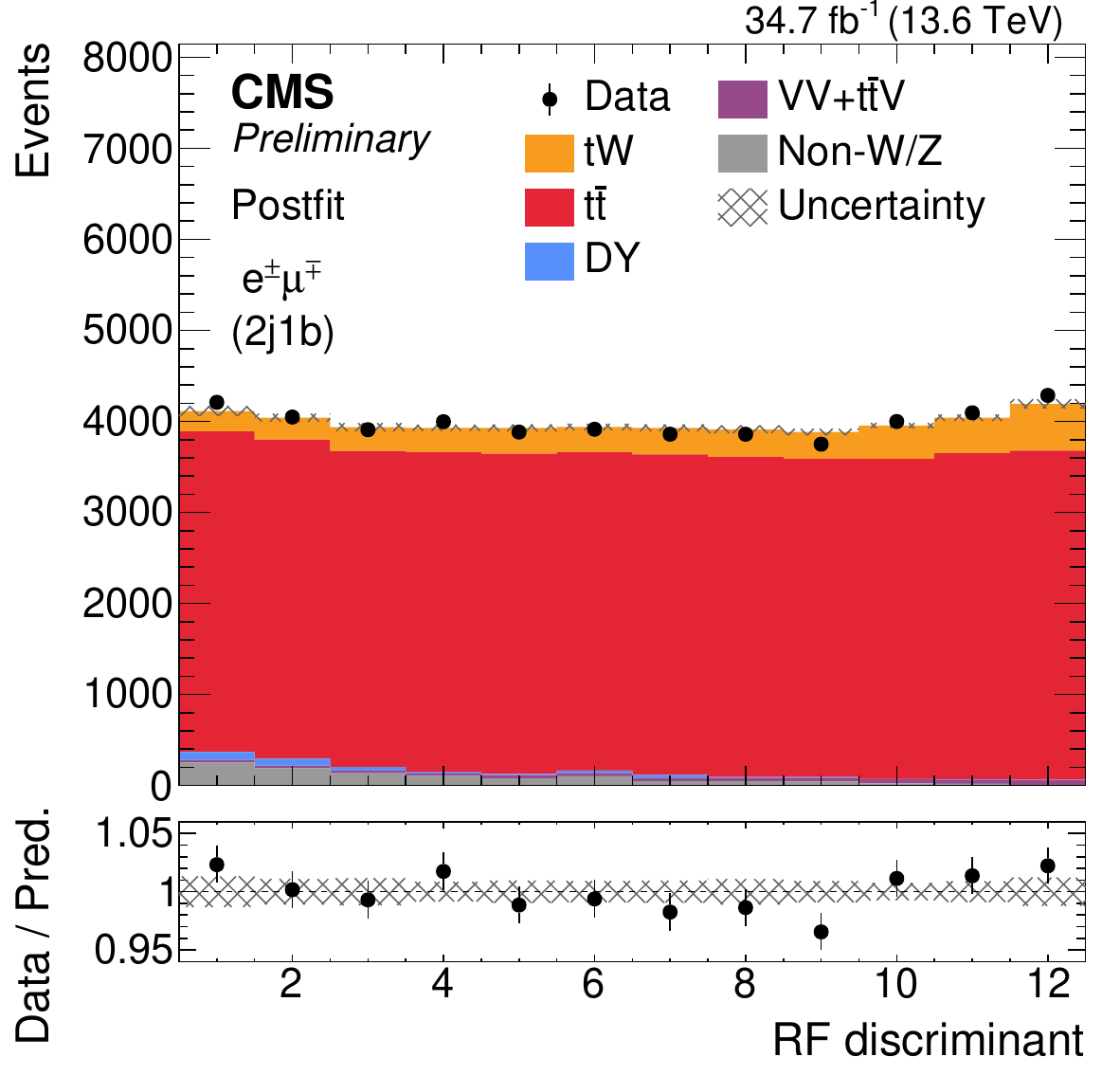}}
\end{minipage}
\hfill
\begin{minipage}{0.328\linewidth}
\centerline{\includegraphics[width=\linewidth]{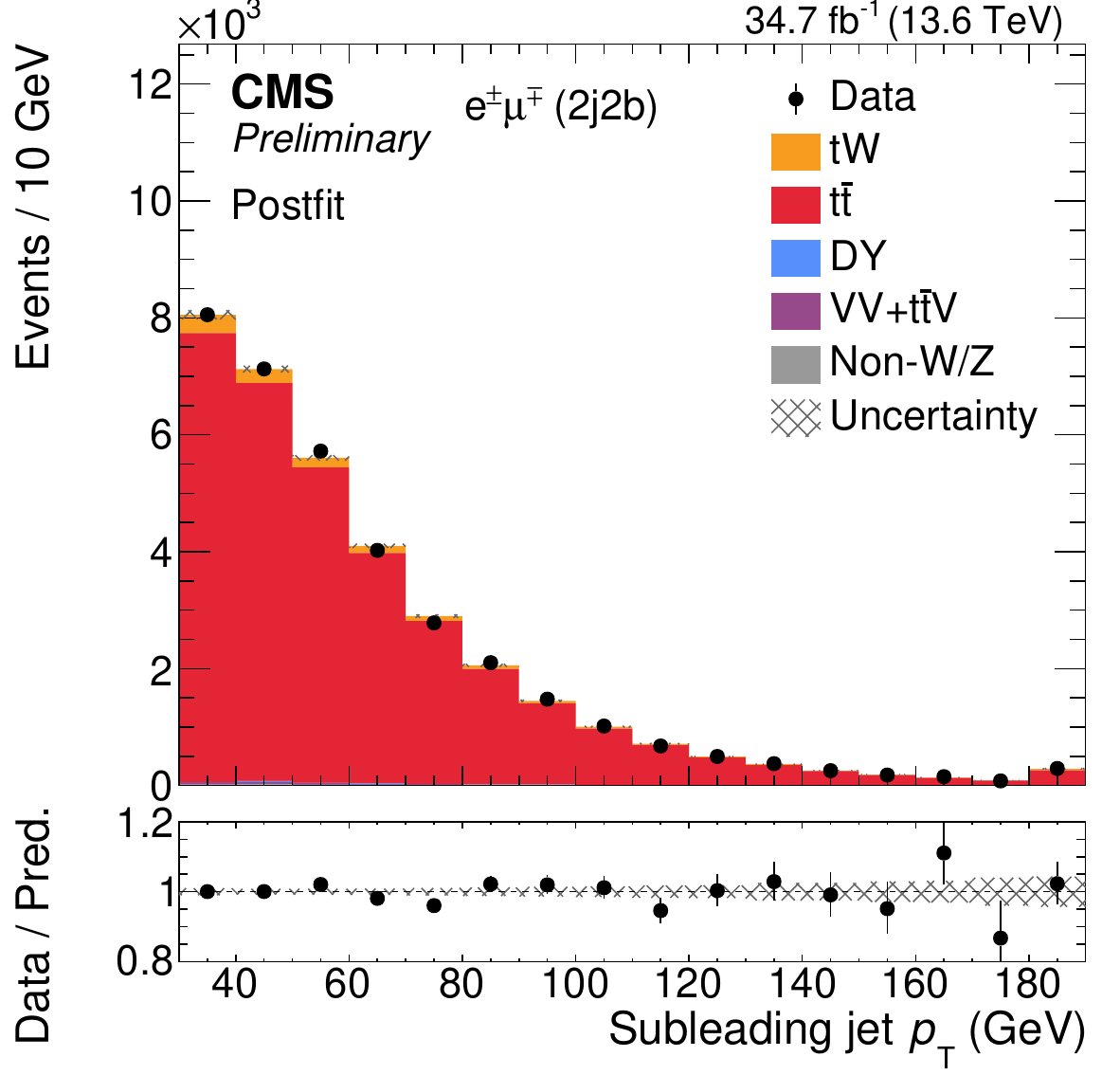}}
\end{minipage}
\caption[]{The distributions of the RF outputs for events in the 1j1b (left) and 2j1b (centre) regions, and the subleading jet $\pt$ for the 2j2b region (right)~\cite{CMS:2024jmu}. The data (points) and the predictions (filled histograms) after the maximum likelihood fit are shown.}
\label{fig:PostFitDistributions}
\end{figure}

The measured inclusive cross section is obtained by performing a maximum likelihood fit to the RF discriminants in the 1j1b and 2j1b regions and the subleading jet $\pt$ distribution in the 2j2b region. These distributions, after the result from the fit is applied, are shown in Fig.~\ref{fig:PostFitDistributions}. The binning of the RF output distribution is chosen such that each bin contains about the same number of \ttbar $ $ events. This avoids the presence of low-statistic bins in the background estimation, which would erroneously constrain the systematic uncertainties. 

We measure a tW inclusive cross section of
\begin{equation}
    \sigma^{\mathrm{exp}}_{\mathrm{tW}} = 84.1 \pm 2.1 (\mathrm{stat})\,^{+9.8}_{-10.2} (\mathrm{syst}) \pm 3.3 (\mathrm{lumi}) \ \mathrm{pb,}
\end{equation}
that is consistent with the SM expectation.

\section{Differential cross section measurements}
Differential cross section measurements provide results free from detector effects to be directly compared with theoretical predictions. Unfolding techniques are used to determine the distributions without the detector effects. In this analysis, unfolding is performed from detector level to particle level using \texttt{TUnfold}~\cite{Schmitt:2012kp}. These measurements are performed in the 1j1b region vetoing events with low energy jets (loose jets). 

\begin{figure}[htpb!]
\begin{minipage}{0.328\linewidth}
\centerline{\includegraphics[width=\linewidth]{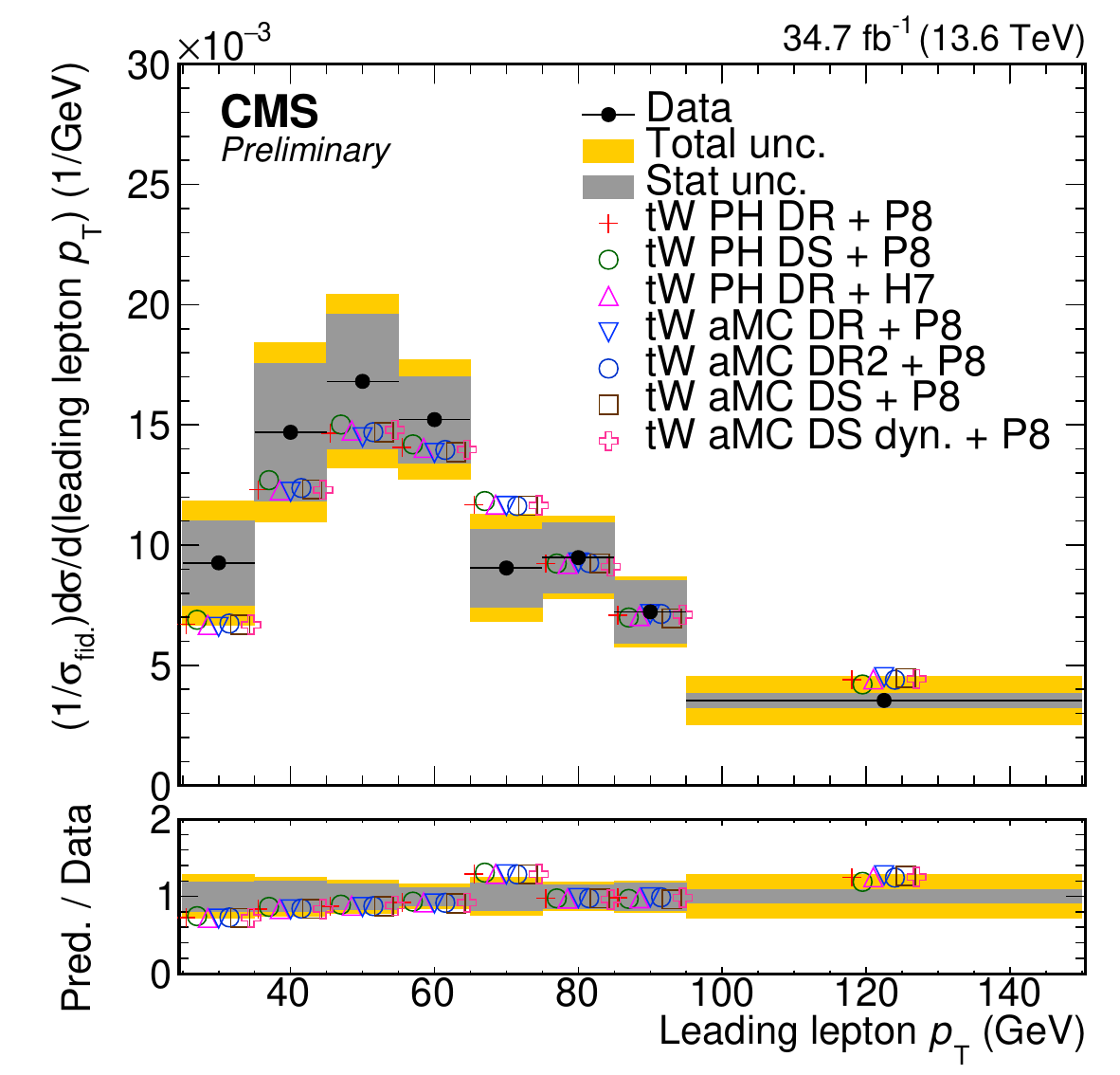}}
\end{minipage}
\hfill
\begin{minipage}{0.328\linewidth}
\centerline{\includegraphics[width=\linewidth]{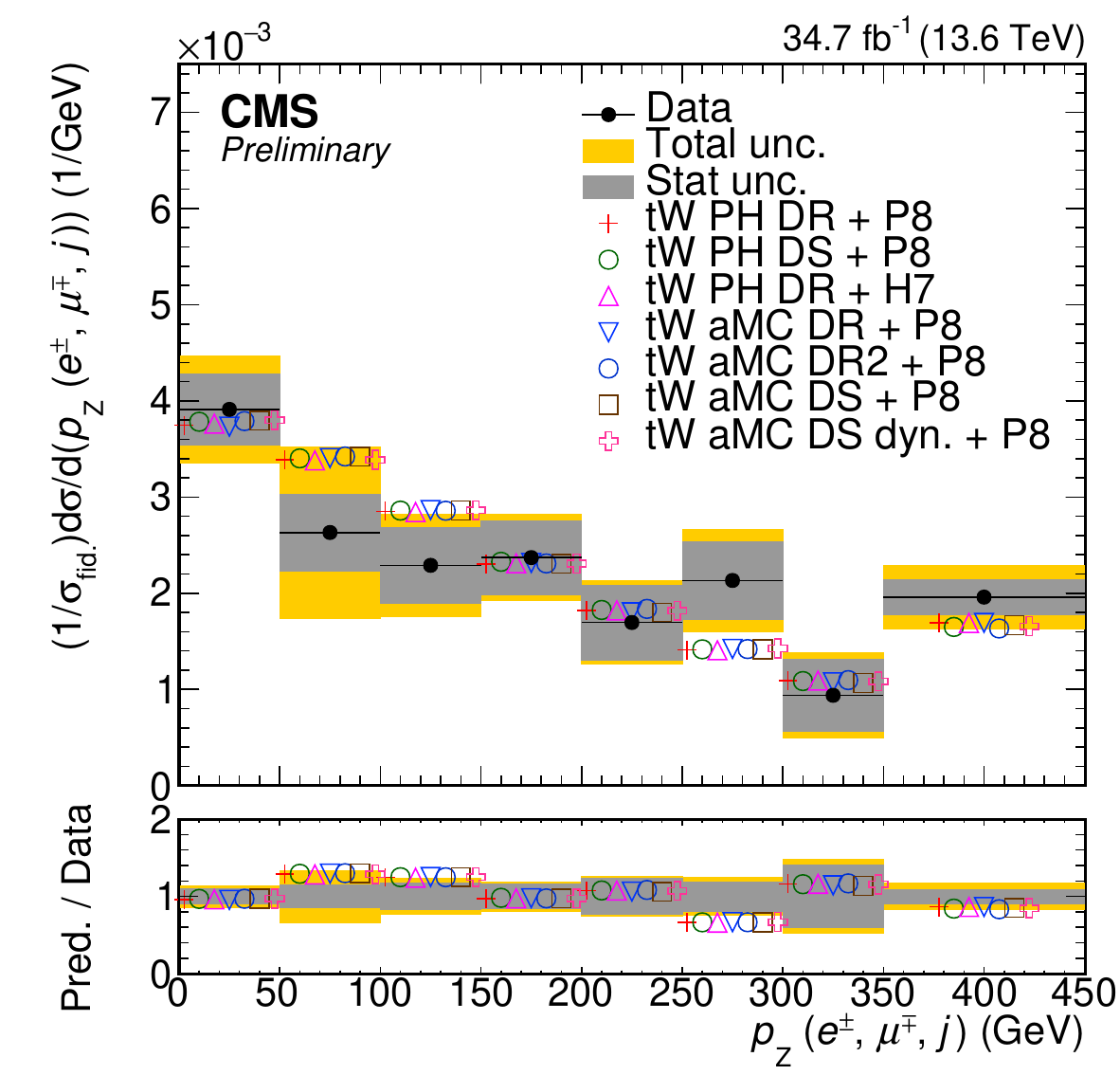}}
\end{minipage}
\hfill
\begin{minipage}{0.328\linewidth}
\centerline{\includegraphics[width=\linewidth]{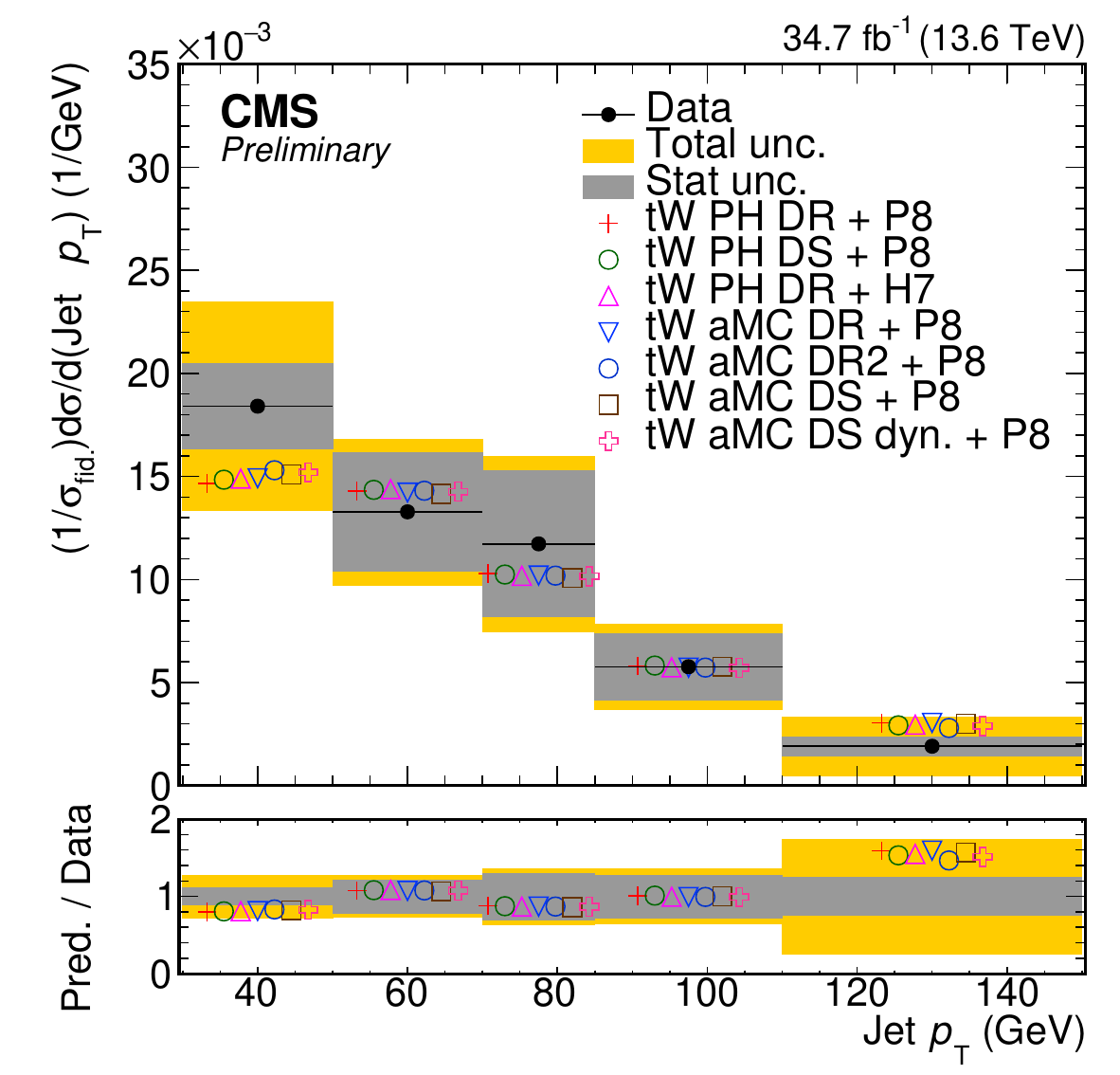}}
\end{minipage}
\begin{minipage}{0.328\linewidth}
\centerline{\includegraphics[width=\linewidth]{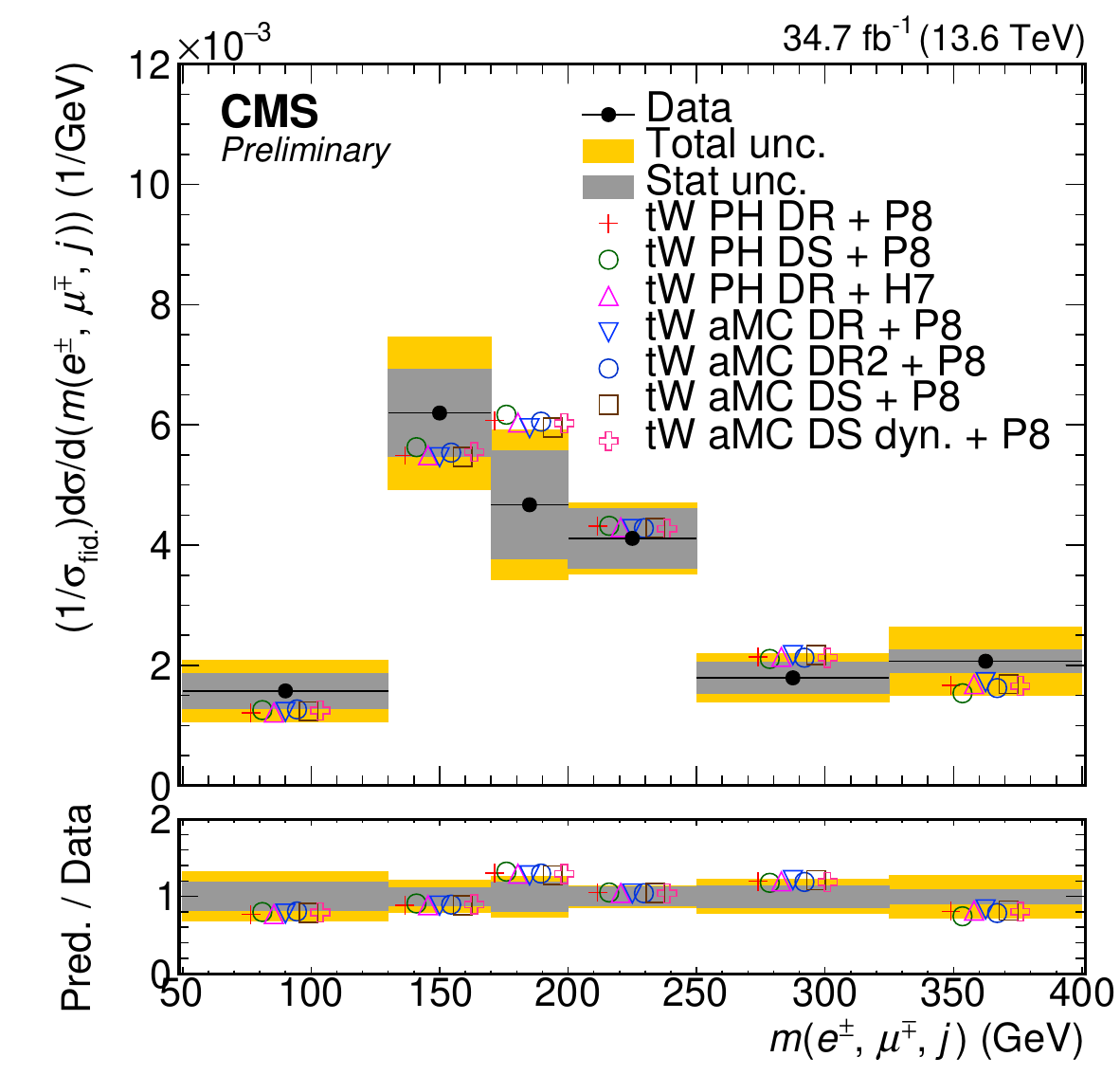}}
\end{minipage}
\hfill
\begin{minipage}{0.328\linewidth}
\centerline{\includegraphics[width=\linewidth]{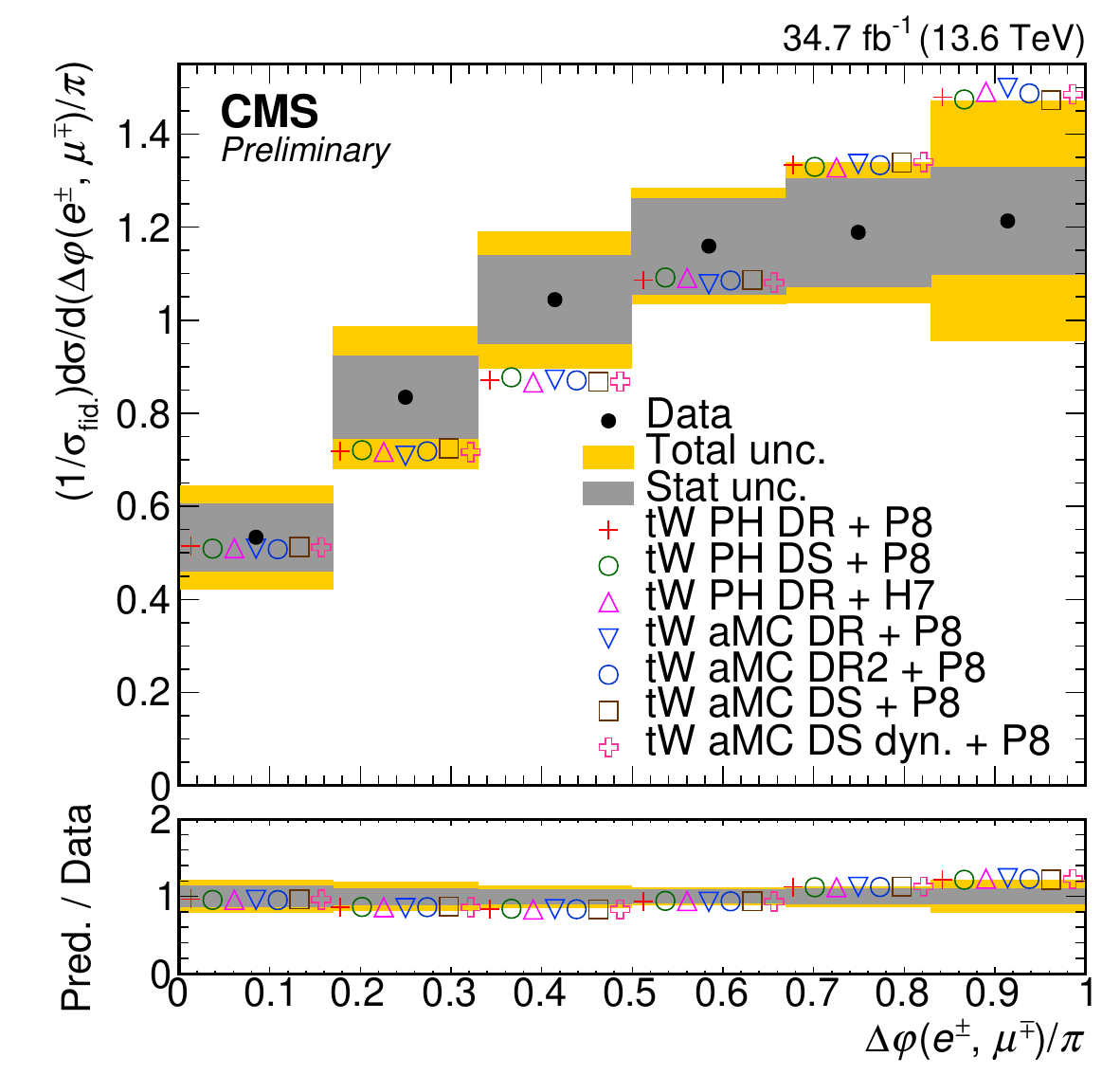}}
\end{minipage}
\hfill
\begin{minipage}{0.328\linewidth}
\centerline{\includegraphics[width=\linewidth]{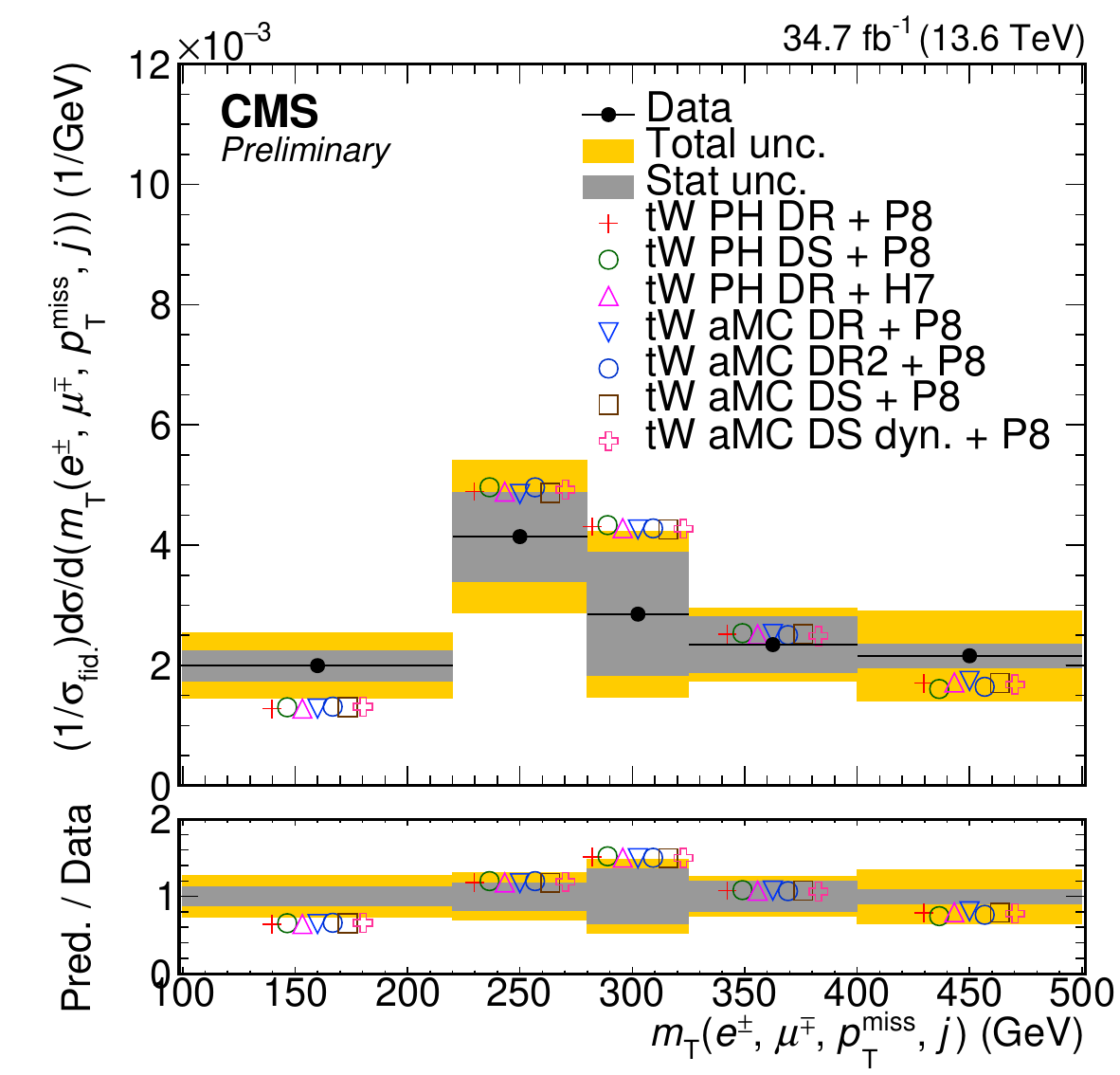}}
\end{minipage}
\caption[]{Normalised differential tW production cross section as functions of the $\pt$ of the leading lepton (upper left), $p_z(e^\pm,\mu^\mp,\mathrm{j})$ (upper centre), $\pt$ of the jet (upper right), $m(e^\pm,\mu^\mp,\mathrm{j})$ (lower left), $\Delta\varphi(e^\pm,\mu^\mp)$ (lower centre), and $m_{\mathrm{T}}(e^\pm,\mu^\mp,\mathrm{j},p_{\mathrm{T}}^{\mathrm{miss}})$ (lower right)~\cite{CMS:2024jmu}.}
\label{fig:particlefidbin1}
\end{figure}

The tW differential cross sections, normalised to the total fiducial cross section $\sigma_{\mathrm{fid.}}$ and bin width, are shown in Fig.~\ref{fig:particlefidbin1} for the data and the simulation predictions. All methods to treat the interference between tW and \ttbar, DR, DR2, DS, and DS with a dynamic factor, show small differences among them. This is also true for the DR predictions interfaced with \texttt{HERWIG}7. The uncertainties, roughly 30--40\% in most cases, depending on the distributions and bins, are dominated by the statistical uncertainties. Good agreement between data and all predictions in the six observables is observed.

\section{Summary}
The first inclusive and normalised differential cross sections for the production of a top quark in association with a W boson are measured in proton-proton collisions at 13.6 TeV. The measured inclusive cross section is in agreement with the prediction. The differential cross section measurements are performed as a function of six kinematical observables of the events. Good agreement between the measurements and the predictions from the different event generators is found.

\section*{References}
\bibliography{AlejandroSotoRodriguez}

\end{document}